# Very strong chalcogen bonding: Is oxygen in molecules capable of forming it? A First-Principles Perspective

Pradeep R. Varadwaj [1]*, Arpita Varadwaj[1,2], Helder M. Marques[3]

[1] Department of Chemical System Engineering, School of Engineering, The University of Tokyo 7-3-1, Tokyo 113-8656, Japan
[2] The National Institute of Advanced Industrial Science and Technology (AIST), Tsukuba 305-8560, Japan
[3] Molecular Sciences Institute, School of Chemistry, University of the Witwatersrand, 1 Jan Smuts Avenue, Braamfontein, Johannesburg 2050, South Africa.

* Corresponding authors address: pradeep@t.okayama-u.ac.jp

Abstract

There are views prevalent in the noncovalent chemistry literature that i) the O atom in molecules cannot form a chalcogen bond, and ii) if formed, this bond is very weak. We have shown here that these views are not necessarily true since the attractive energy between the oxygen atom of some molecules and several electron-rich anionic bases examined in a series of 34 ion-molecule complexes varied from the weak (ca –2.30 kcal mol$^{-1}$) to the ultra-strong (–90.10 kcal mol$^{-1}$). The [MP2 /aug-cc-pVTZ] binding energies for several of these complexes were found to be comparable to or significantly larger than that of the well-known hydrogen bond complex [FH⋯F]$^−$ (~ 40 kcal mol$^{-1}$). The nature of the intermolecular interactions was examined using the quantum theory of atoms in molecules, second-order natural bond orbital and symmetric adaptive perturbation theory energy decomposition analyses. It was found that many of these interactions comprise mixed bonding character (ionic and covalent), especially manifest in the moderate to strongly bound complexes. All these can be explained by an $n$ (lone-pair bonding orbital) → σ* (anti-bonding orbital) donor-acceptor charge transfer delocalization. This study, therefore, demonstrates that the covalently bound oxygen atom in molecules can have a significant ability to act as an unusually strong chalcogen bond donor.

1. **Introduction**

Chalconide (Ch) oxygen is generally regarded as the least polarizable element in Group 16 of the periodic table.[1,2] Because of this, several studies have demonstrated that it does not form a chalcogen bond.[3-7] (A chalcogen bond is formed when there is evidence of a positive site on the Ch atom in a molecule that interacts attractively with a negative site on a Lewis base in another molecule.[8-15]) The argument is that the oxygen atom in molecules is often negative and therefore does not feature a positive σ-hole on its electrostatic surface that can attract the negative site on a base.[3-15] (A σ-hole[16,17] is an electron density deficient region on the surface of the atom Ch along the outer portion of the R–Ch bond axis, where R is the remainder part of the molecule.[18]) While this may indeed be so in many cases (such as $H_2O$ and $H_2CO$),[8,9] it should not always be taken for granted.

We have recently shown that when the oxygen atom is covalently bonded to an electron-withdrawing group X (X = F, Cl, Br) and –CN, the group draws the electron density to the bonding region and generates a weak to a moderately strong electron density deficient-region (σ-hole) on the surface of the atom lying opposite to the bond.[8,9] The interaction energy (also called the binding energy) of putative 1:1 complexes formed by the O atom and the Lewis bases was found to be small (< 3 kcal mol$^{-1}$).[8,9] This has led to the interpretation that the aforementioned complexes may be regarded as being formed by van der Waals or weak interaction. These studies on O-centered chalcogen bonding have been recognized by others, both experimentally and theoretically.[19,20] However, some have claimed that these complexes may not involve true "chalcogen bonding" since the intermolecular interactions in them are weakly bound and "chalcogen bonding" should be regarded as an "electrostatically driven" interaction – reminiscent of a debate that is very common in the area of halogen bonding.[21,22] We counter this view by noting that there is no hard and fast rule in which a "weakly bound" or "van der Waals" complex cannot be regarded as truly "chalcogen bonded". The minimal criterion for the recognition of a "Type-II chalcogen bond"[11,23] is that there must be an attraction between the electrophilic region on the Ch atom and a negative site, and that the angle of approach of the electrophile should be such that ∠X···Ch–R = 140–180°.[8-15]

We add further that the importance of weak interactions should not be underestimated.[24,25] They appear in many different flavors,[25-32] yet an understanding of their physical and chemical behavior in chemical systems to date is not complete. They are important factors, for example, not only in the theoretical and experimental design of drug[33,34] and polymer network structures,[35-37] but also for developments in the fields of crystal engineering[38] and molecular recognition.[24,39] The hydrogen bond between two $H_2O$ molecules is also weak, and is never stronger than about a twentieth the strength of the O–H covalent bond. Such bonds are structure determining, and have profound significance in fields as diverse as biology and materials science.[40] The energy of a van der Waals interaction is very weak, only about 1 kcal mol$^{-1}$, comparable with the average kinetic energy of a molecule in solution (approximately 0.4 kcal mol$^{-1}$). It is significant only when many of them are combined so they contribute to the overall structure of a chemical system (as in interactions of complementary surfaces).[41,42] Accordingly, and based on their energy of stability preferences, intermolecular interactions have been classified as van der Waals (energy < 1 kcal mol$^{-1}$),[43] weak (1–4 kcal mol$^{-1}$),[43,44] moderate (4–15 kcal mol$^{-1}$),[45] strong (15–40 kcal mol$^{-1}$),[44-46] very strong (40–60 kcal mol$^{-1}$)[47] and ultra-strong (>> 60 kcal mol$^{-1}$).[47-51] The first four have been recognized in many systems, while the last two have been identified in singly- and doubly charge-assisted composite systems, respectively.[48,51-53]

Whereas thousands of studies have been reported centering discussion on the chemical physics and physical chemistry of halogen bonding and chalcogen bonding interactions between molecules containing heavy atom donors, the exploration of the chemistry of O-center chalcogen bonding from a theoretical modeling perspective is very limited. To this end, we report here an investigation of the structure, energy, electronic, orbital, and topological properties of 34 ion-molecule complexes formed by the attractive engagement between the positive sites (positive σ-holes) on the O atoms of some O-containing molecules and a series of anions. Analogous molecule-anion complexes, formed by hydrogen- and halogen-bonds are well known and hundreds of structures featuring these have been deposited in the Cambridge Structure Database.[54] Kumar and coworkers, for instance, have examined the robustness of stable benzylic selenocynates for halide ion recognition in the solid-state and in solution.[55] Their XRD analysis of various cocrystals reveals that the NCSe···X$^-$ systems are driven by structurally important Se···X$^-$ (X = Cl, Br, I) chalcogen bonds. Similar studies have been reported by ohers.[56,57] Similarly, Galmés and coworkers have recently performed a combined Cambridge Structural Database and theoretical DFT study of charge assisted chalcogen bonds involving sulfonium, selenonium, and telluronium cations in which divalent chalcogen atoms typically have up to two σ-holes and form up to two chalcogen bonds; the same holds for tetravalent chalcogens which adopt a seesaw arrangement.[58] Analogous studies have been reported elsewhere.[59] However, the complexes examined in this study are uncommon; they are promoted by O-centered chalcogen bonding. The results of the *ab initio* first-principles MP2 method[60] show that the binding energy of many of these complexes can be unusually high, comparable to or greater than that of various halogen- and hydrogen-bonded systems already reported in the noncovalent chemistry literature. In addition, we used the results of symmetry adapted perturbation theory (SAPT),[61,62] the second-order perturbative estimates of 'donor-acceptor' (bond-antibond) interaction energies in the natural bond orbital (NBO) basis,[63] and the quantum theory atoms in molecules (QTAIM)[64] to show that the intermolecular interactions responsible for the formation of the ion-molecule complexes contain appreciable covalent character. Of course, this adds to their inherent ionic character, which can well be rationalized by the Coulomb's law.

2. **Chemical model systems and computational details**

The binary complexes of $OX_2$ (X = F, Cl, Br, CN) with the anions $A^-$ (A = F, Cl, Br, CN, $Br_3$, SCN, NCO, $NO_3$) were fully energy minimized using MP2 (fc), in conjunction with the aug-cc-pVTZ basis set; the reliability of this and other theoretical approaches to study noncovalent interactions has been discussed elsewhere.[8,9,65-67] The calculation of the Hessian second derivative of the energy with respect to the fixed nuclear coordinates of the atom was performed for all cases to ensure that a true minimum was found; all eigenvalues were found to be positive, and the structures reported here are not transition states. All calculations were formed using Gaussian 09.[68]

The nature of electrostatic surface of each $OX_2$ molecule was examined using the popular molecular electrostatic surface potential (MESP) approach.[69] As has been done elsewhere,[8,69] the 0.001 a.u. isodensity envelopes of these molecules were used on which to compute the potential. The local maxima and minima of potential, often referred to as $V_{S,max}$ and $V_{S,min}$, respectively, were used to identify the positive and negative regions, respectively. It should be kept in mind that the sign of $V_{S,max}$ and $V_{S,min}$ is not always positive or always negative. The positive/negative sign associated with these two properties depends on the nature of the nucleophilicity/electrophilicity of a specific region on an atom or fragment in a molecule. Nevertheless, when $V_{S,max} > 0$ (or $V_{S,max} < 0$) on atom Ch along the outer extension of the R–Ch bond, it identifies a positive (or a negative) σ-hole.[17] Similarly, when $V_{S,min} > 0$ (or $V_{S,min} < 0$) on a specific region, it signifies an electrophilic (or nucleophilic) site. Both $V_{S,max}$ and $V_{S,min}$ were calculated using Multiwfn,[70] and the MESP plots were generated using AIMAll[71] with the wavefunctions generated using the [MP2/aug-cc-pVTZ] geometries. The counterpoise method of Boys and

Bernardi was invoked to account for the effect of Basis Set Superposition Error (BSSE) on energy.[72]

QTAIM calculations were performed with [MP2/aug-cc-pVTZ] – a theory that relies on the zero-flux boundary condition to partition atomic domains in real space.[64,73,74] The typical topological properties such as the gradient paths, bond paths, bond critical points (bcps) of the charge density ($\rho_b$), the Laplacian of the charge density ($\nabla^2\rho_b$) and the total energy density ($H_b$) were evaluated. In addition, the delocalization indices, $\delta$, between various atom-atom pairs were evaluated for each system to gain insight into the covalent nature of the various bonding interactions involved.[75,76]

We examined the nature of the charge transfer delocalization energies $E^2$ between "filled" (donor) Lewis-type NBOs and "empty" (acceptor) non-Lewis type NBOs in several complexes – all within the second-order framework of NBO analysis (Eqt. 1).[63] In Eqn. 1, $q_i$ is the donor orbital occupancy, $\varepsilon_i$ and $\varepsilon_j$ are diagonal elements (orbital energies) associated with each donor NBO ($i$) and acceptor NBO ($j$), respectively, and $F(i,j)$ is the off-diagonal NBO Fock matrix element. These calculations were carried out within the Hartree–Fock (HF) level theory using Gaussian 09's NBO Version 3.1.[63]

$$E^2 = \Delta E_{ij} = q_i \frac{F(i,j)^2}{\varepsilon_i - \varepsilon_j} \tag{1}$$

SAPT-based energy decomposition analysis (EDA) was performed within the framework of density functional theory. The [MP2/aug-cc-pVTZ] geometries of the complexes were used. The SAPT theory dissects the interaction energy $E_{int}$(SAPT) of a complex into four major components: electrostatic ($E_{eles}$), exchange ($E_{exch}$), polarization/induction ($E_{ind}$), and dispersion ($E_{disp}$). As such, $E_{int}$(SAPT) is approximated by Eqn. 2, where the subscript, *resp*, indicates that the orbital relaxation effects are included. The term $\delta^{(2)}_{HF}$ takes into account higher-order induction effects and is included in the definition of SAPT terms. The details of each term used in Eqn. 2 are discussed elsewhere.[61,62] The PSI4 code was utilized.[62]

$$E_{int}(SAPT) = E^{(10)}_{eles} + E^{(10)}_{exch} + E^{(20)}_{ind,resp} + E^{(20)}_{exch-ind,resp} + E^{(20)}_{disp} + E^{(20)}_{exch-disp} + \delta^{(2)}_{HF} \qquad (2)$$

### 3. Results and discussion

**3.1 QTAIM topologies and intermolecular geometries**

The QTAIM molecular graphs, showing the possibility of various bonding interactions between atoms in the complex anions examined are given in Fig. 1. Evidence of these interactions come from the presence of well-defined bond paths and (3,–1) bcps between the bonded atomic basins in each binary complex. Such topological signatures are expected when the constituent atoms in the monomers are engaged in the formation of either covalent or noncovalent interactions. The results are consistent with the IUPAC recommendation for identifying the presence of chalcogen bonding.[77]

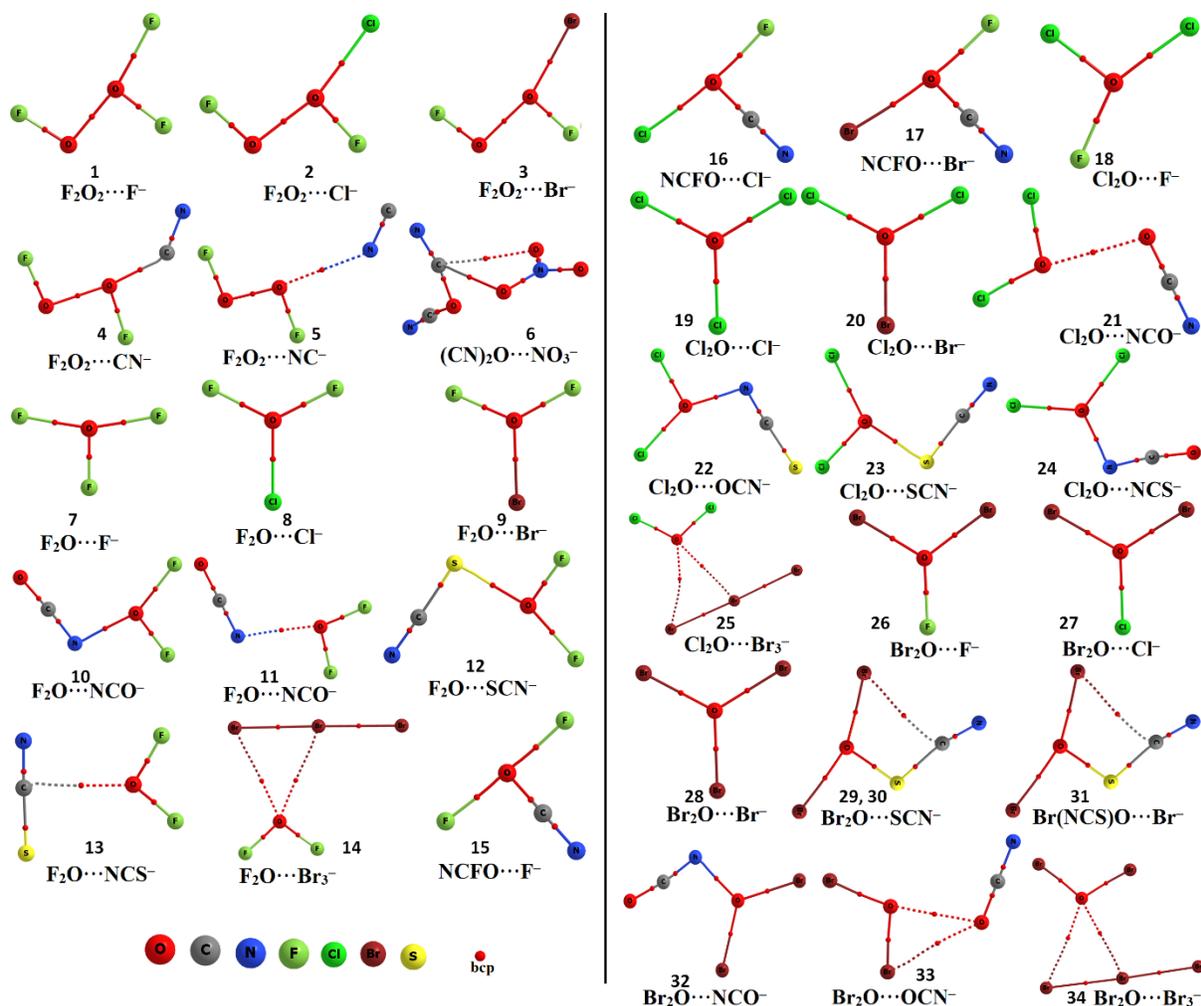

**Fig. 1:** QTAIM molecular graphs of all the O-bonded ion-molecule complexes, obtained with MP2/aug-cc-pVTZ. Atom identity is labeled. The bond paths are shown as solid or dotted lines in atom color and bond critical pints as tiny red spheres between bonded atomic basins.

The values of $\rho_b$ at the bcps between $OX_2$ and anions were found to be between 0.008 and 0.253 a.u. (Table S1 of the Supplementary Information). Except at the bcp of $Br_2O\cdots SCN^-$ (29), the Laplacian of the charge density, $\nabla^2\rho_b$, is large and positive for all complexes, with values varying between 0.020 and 0.416 a.u. Since $\nabla^2\rho_b > 0$ and $\nabla^2\rho_b < 0$ have been regarded as a signature of closed- and open-shell (ionic) interactions,[78] the $\nabla^2\rho_b$ = –0.258 a.u. at the bcp of $Br_2O\cdots SCN^-$ signifies the presence of an open-shell (covalent) interaction.[73,74]

The O⋯A (A = F, Cl, Br, C, N) contacts in the complexes of $O_2F_2$ with the anions $A^-$ (A = F, Cl, Br, CN) (1-5, Fig. 1) exhibit similar closed-shell bonding features ($\nabla^2\rho_b > 0$), yet the angles of approach of the electrophile varies between 162.0 and 170.4° (Table 1). The directionality of the interaction is comparable with that obtained for the OFCN complexes of the anions $A^-$ (A = F, Cl, Br) (15-17), with the A⋯O–C (A = F, Cl, Br) angles ranging between 168.9 and 175.7°. It is comparatively weaker for the complexes of $OF_2$, $OCl_2$ and $OBr_2$ with the same anions (7-9, 18-20 & 26-28), with the A⋯O–X angles commensurate with the range typically found for Type-II interactions (150–180°).[17]

The complexes $F_2O⋯Br_3^-$, $Cl_2O⋯Br_3^-$ and $Br_2O⋯Br_3^-$ illustrated as 14, 25 and 34 in Fig. 1, respectively, display the presence of a bifurcated topology of bonding, in which the O atom in $X_2O$ acts as an electrophile for the $Br_3^-$ anion. Because of the involvement of a secondary interaction in each of these three complexes, the Br⋯O–X angles are significantly non-linear. Similar topologies of bonding are also found for the complexes shown as 6 and 33 (($(CN)_2O⋯NO_3^-$ and $Br_2O⋯OCN^-$, respectively). In 6, the bond path topologies are developed between the $(O)C_\pi$ atom of the $(CN)_2O$ molecule and the two O atoms of the $NO_3^-$ anion. In 33, such topologies are seen between the O (and Br) atom(s) of the $OBr_2$ molecule and the O atom of the $OCN^-$ anion. Since the σ-hole on the O atom in $OBr_2$ is weakly positive ($V_{S,max}$ = +1.6 kcal mol$^{-1}$, Fig. 2), its attractive interaction with the entirely negative O atom in $OCN^-$ results in the formation of the O⋯O Type-II chalcogen bond (∠O⋯O–Br = 166.8°). The other interaction in the same complex is secondary, and is classified as Type-I bonding (∠Br⋯O–C = 139.2°). The combination of both the Type-I and -II interactions provides geometrical stability to the entire ion-molecule complex. Details of the intermolecular distance and directional nature of various A⋯O interactions in other complexes are summarized in Table 1.

**Table 1.** The uncorrected and BSSE-corrected binding energies and selected geometrical properties of the 34 O-bonded complexes, obtained with [MP2/aug-cc-pVTZ].[a]

| No. | Complex | $\Delta E$(MP2) | $\Delta E$(BSSE)(MP2) | Distance | Value | Angle | Value |
|---|---|---|---|---|---|---|---|
| 1 | $F_2O_2\cdots F^-$ | -23.27 | -22.23 | F$\cdots$O | 1.849 | $\angle$F$\cdots$O–O | 167.7 |
| 2 | $F_2O_2\cdots Cl^-$ | -23.95 | -22.78 | Cl$\cdots$O | 2.196 | $\angle$Cl$\cdots$O–O | 163.1 |
| 3 | $F_2O_2\cdots Br^-$ | -28.42 | -26.14 | Br$\cdots$O | 2.297 | $\angle$Br$\cdots$O–O | 162.0 |
| 4 | $F_2O_2\cdots CN^-$ | -27.02 | -25.52 | C$\cdots$O | 1.815 | $\angle$C$\cdots$O–O | 168.1 |
| 5 | $F_2O_2\cdots NC^-$ | -4.32 | -3.82 | N$\cdots$O | 2.654 | $\angle$N$\cdots$O–O | 170.4 |
| 6 | $(CN)_2O\cdots NO_3^-$ | -19.62 | -18.36 | O$\cdots$O | 2.600 | $\angle$F$\cdots$O–C | 177.3 |
| 7 | $F_2O\cdots F^-$ | -38.96 | -37.74 | F$\cdots$O | 1.742 | $\angle$F$\cdots$O–F | 159.5 |
| 8 | $F_2O\cdots Cl^-$ | -55.81 | -54.10 | Cl$\cdots$O | 1.961 | $\angle$Cl$\cdots$O–F | 117.2 |
| 9 | $F_2O\cdots Br^-$ | -57.97 | -54.93 | Br$\cdots$O | 2.082 | $\angle$Br$\cdots$O–F | 117.1 |
| 10 | $F_2O\cdots NCO^-$ | -33.28 | -31.84 | N$\cdots$O | 1.894 | $\angle$N$\cdots$O–F | 153.8 |
| 11 | $F_2O\cdots NCO^-$ | -6.66 | -6.08 | N$\cdots$O | 2.501 | $\angle$N$\cdots$O–F | 162.7 |
| 12 | $F_2O\cdots SCN^-$ | -42.49 | -40.90 | S$\cdots$O | 2.138 | $\angle$S$\cdots$O–F | 161.1 |
| 13 | $F_2O\cdots NCS^-$ | -5.48 | -4.89 | N$\cdots$O | 2.999 | $\angle$N$\cdots$O–F | 161.0 |
|  |  |  |  | S$\cdots$O | 3.188 | $\angle$S$\cdots$O–F | 152.2 |
| 14 | $F_2O\cdots Br_3^-$ | -4.88 | -3.65 | Br$\cdots$O | 3.120 | $\angle$Br$\cdots$O–F | 154.6 |
|  |  |  |  | Br$\cdots$O | 3.154 | $\angle$Br$\cdots$O–F | 151.5 |
| 15 | NCFO$\cdots$F$^-$ | -52.2 | -50.86 | F$\cdots$O | 1.797 | $\angle$F$\cdots$O–F | 175.7 |
| 16 | NCFO$\cdots$Cl$^-$ | -50.26 | -48.74 | Cl$\cdots$O | 2.132 | $\angle$Cl$\cdots$O–F | 170.1 |
| 17 | NCFO$\cdots$Br$^-$ | -54.42 | -51.46 | Br$\cdots$O | 2.241 | $\angle$Br$\cdots$O–F | 168.9 |
| 18 | $Cl_2O\cdots F^-$ | -25.64 | -24.42 | F$\cdots$O | 1.809 | $\angle$F$\cdots$O–Cl | 149.9 |
| 19 | $Cl_2O\cdots Cl^-$ | -39.29 | -37.56 | Cl$\cdots$O | 1.974 | $\angle$Cl$\cdots$O–Cl | 120.0 |
| 20 | $Cl_2O\cdots Br^-$ | -45.85 | -42.57 | Br$\cdots$O | 2.062 | $\angle$Br$\cdots$O–Cl | 118.5 |
| 21 | $Cl_2O\cdots NCO^-$ | -35.83 | -33.97 | N$\cdots$O | 1.801 | $\angle$N$\cdots$O–Cl | 120.5 |
| 22 | $Cl_2O\cdots OCN^-$ | -2.93 | -2.30 | O$\cdots$O | 2.653 | $\angle$Br$\cdots$O–Cl | 166.1 |
| 23 | $Cl_2O\cdots SCN^-$ | -36.06 | -34.27 | S$\cdots$O | 2.110 | $\angle$S$\cdots$O–Cl | 149.6 |
| 24 | $Cl_2O\cdots NCS^-$ | -13.23 | -11.72 | N$\cdots$O | 1.952 | $\angle$N$\cdots$O–Cl | 148.9 |
| 25 | $Cl_2O\cdots Br_3^-$ | -4.41 | -2.80 | Br$\cdots$O | 3.328 | $\angle$Br$\cdots$O–Cl | 137.2 |
|  |  |  |  | Br$\cdots$O | 3.084 | $\angle$Br$\cdots$O–Cl | 159.5 |
| 26 | $Br_2O\cdots F^-$ | -24.61 | -22.93 | F$\cdots$O | 1.771 | $\angle$F$\cdots$O–Br | 124.0 |
| 27 | $Br_2O\cdots Cl^-$ | -32.9 | -30.77 | Cl$\cdots$O | 2.002 | $\angle$Cl$\cdots$O–Br | 121.2 |
| 28 | $Br_2O\cdots Br^-$ | -40.37 | -36.64 | Br$\cdots$O | 2.075 | $\angle$Br$\cdots$O–Br | 120.0 |
| 29 | $Br_2O\cdots SCN^-$ | -94.61 | -90.07 | S$\cdots$O | 1.742 | $\angle$S$\cdots$O–Br | 91.2 |
| 30 | $Br_2O\cdots NCS^-$ | -94.61 | -90.07 | S$\cdots$O | 1.742 | $\angle$S$\cdots$O–Br | 91.2 |
| 31 | Br(NCS)O$\cdots$Br$^-$ | -93.11 | -89.54 | Br$\cdots$O | 2.242 | $\angle$Br$\cdots$O–Br | 156.8 |
| 32 | $Br_2O\cdots NCO^-$ | -30.20 | -27.50 | N$\cdots$O | 1.829 | $\angle$N$\cdots$O–Cl | 121.8 |
| 33 | $Br_2O\cdots OCN^-$ | -3.65 | -2.36 | O$\cdots$O | 2.597 | $\angle$O$\cdots$O–Br | 166.8 |
|  |  |  |  | O$\cdots$Br | 2.987 | $\angle$Br$\cdots$O–C | 139.2 |
| 34 | $Br_2O\cdots Br_3^-$ | -5.51 | -3.03 | Br$\cdots$O | 3.177 | $\angle$Br$\cdots$O–Br | 141.0 |
|  |  |  |  | Br$\cdots$O | 3.019 | $\angle$Br$\cdots$O–Br | 155.9 |

[a] Energies in kcal mol$^{-1}$. Selected bond lengths and bond angles in Å and degrees, respectively.

The total energy density, $H_b$, is sum of the "gradient" kinetic energy density, $G_b$, and the potential energy density, $V_b$, ($H_b = G_b + V_b$). From Table S1, it is clear that $H_b$ is negative for several complexes, implying that they conceive medium-to-strongly bound interactions. For instance, it is most negative for the S⋯O bond ($H_b$ = –0.166 a.u.) in Br$_2$O⋯SCN$^−$. Both this and the $\nabla^2\rho_b$ of –0.258 a.u. at the S⋯O bcp signify the formation of a covalent bond between the S and O atoms in the complex. This conclusion is consistent with the recommendation of Cremer and Kraka,[78] who have suggested that if $V_b$ is dominant over $G_b$ at the bcps (hence, $H_b < 0$), this indicates that the region has a "charge density concentration". On the other hand, and for the weakly bound interactions in the complexes shown in 5-6, 11, 13-14, 22, 25, 29-30 and 33-34 of Fig. 1, the $H_b$ values are all positive. This is not unexpected since the intermolecular distances in these complexes are relatively longer (Table 1) and the nature of these interactions are closed-shell type. This occurs when there is an appreciable depletion in the charge density at the bcps of bonded atomic basins. A similar insight into the ionic and covalent nature of the aforesaid interactions can be gained from the ratio $-V_b/G_b$ (values not shown).[79,80]

**3.2 Analysis of the nature of electron-pair delocalization between atomic basins**

The delocalization index $\delta$, a measure of the bond order, is calculated between various atom-atom pairs constituting each complex. The largest value of $\delta$ was found to be 0.896 for the O and S atom-atom pair in the complex Br$_2$O⋯SCN$^−$ (29); this means that approximately one electron is shared between these atomic basins. This leads to the suggestion that there is indeed formation of a covalent bond, and the entire complex system might be regarded as a single molecular cation.

The A⋯O (X = F, Cl, Br, S, N) atom-atom pairs formed between the O-centered monomers and other anions are also accompanied by reasonably large $\delta$ values (Table

S1). Similar $\delta$ values have been reported for coordination bonds. For example, $\delta$ for the Ru—C, Ru—H and Ru—Ru bonds in the *N*-heterocyclic carbene triruthenium clusters [Ru$_3$($\mu$-H)$_2$($\mu_3$-MeImCH)(CO)$_9$](Me$_2$Im = 1,3-dimethylimidazol-2-ylidene) were reported as 0.757–0.764, 0.474 and 0.458, respectively.[81] Similarly, the δ for the Co—Co and Fe—Fe bonds were 0.47 and 0.40 for Co$_2$(CO)$_8$ and Fe$_2$(CO)$_8$, respectively.[82] Clearly, the combined signature ($H_b$ < 0 and $\delta$ >> 0.4) unequivocally indicates that the A···O intermolecular interactions in 25 complexes of the entire series have significant covalency. On the other hand, the $\delta$ values are significantly smaller for the O···O, O···N, O···F, O···Cl and O···Br atom-atom pairs of the complexes shown in 5-6, 11, 13-14, 22, 25, 29-30 and 33-34.

**3.3 Binding energies**

The uncorrected binding energy, $\Delta E$, was calculated using the supermolecular procedure proposed by Pople.[83] That is, $\Delta E$ is the difference between the sum of the total electronic energies of two monomer species (OX$_2$ and A$^-$) and the total electronic energy of the complex X$_2$O···A$^-$ (A = F, Cl, Br, CN, Br$_3$, SCN, NCO, NO$_3$). We found that the most stable complex Br$_2$O···SCN$^-$ has a $\Delta E$ of –90.05 kcal mol$^{-1}$, while the least stable complex Cl$_2$O···OCN$^-$ has a $\Delta E$ of –2.30 kcal mol$^{-1}$ Table 1).

The $\Delta E$ of F$_2$O···CN$^-$ (–25.52 kcal mol$^{-1}$) is significantly larger than that of F$_2$O···NC$^-$ (–4.32 kcal mol$^{-1}$), indicating that the C-end of the CN$^-$ anion is significantly more nucleophilic that the N-end. Similarly, the $\Delta E$ of the complex X$_2$O···NCO$^-$ (X = F, Cl, Br) is significantly larger than that of the corresponding X$_2$O···OCN$^-$ complex, showing that the O in OCN$^-$ is a poorer nucleophile than N. This feature is consistent with SCN$^-$ as well, where the X$_2$O···SCN$^-$ (X = F, Cl) complexes are markedly stronger than the corresponding X$_2$O···NCS$^-$ complexes. In the case of Br$_2$O···SCN$^-$ and Br$_2$O···NCS$^-$, the energy minimization procedure changed the geometry of the latter complex into that of the former, with a $\Delta E$ of –94.61 kcal mol$^{-1}$. This suggests that the Columbic repulsion between the N and O atoms of the interacting monomers is predominant, which prevents the N-end of NCS$^-$ atttracting the O-end of Br$_2$O, revealing that S is a better nucleophile (hence, more reactive) than N. The complex Br(NCS)O⋯Br$^-$ (31) is equivalent to Br$_2$O⋯SCN$^-$ (29-30). However, a $\Delta E$ of –93.11 kcal mol$^{-1}$ for the O⋯Br bond in this system, given that Br(NCS)O and Br$^-$ are the two monomers of the system, reveals a potentially covalently bound interaction. Although the complex systems 29 and 30 represent the same Br$_2$O⋯SCN$^-$, the system is numbered differently to show that no matter which end of the SCN$^-$ ion is bonded with Br$_2$O the $\Delta E$ between Br$_2$O and SCN$^-$ is the same (Table 1) given that the total electronic energies of these two isolated systems were used for the calculation of $\Delta E$. The effect of basis set superposition error (BSSE) on $\Delta E$ is not negligible for all systems examined, with BSSE values ranging from –4.54 kcal mol$^{-1}$ for the complex Br$_2$O⋯SCN$^-$ (29-30) to –0.50 kcal mol$^{-1}$ for the complex F$_2$O⋯NC$^-$ (5).

The $\Delta E$(BSSE) values for the complexes of X$_2$O⋯A$^-$ and NCFO⋯A$^-$ (A = F, Cl, Br) are comparable with those widely recognized to arise from charge-assisted hydrogen-bonded complexes. For example, the energy of intermolecular interaction for HF⋯F$^-$ was reported to be –38.6 kcal mol$^{-1}$.[43,44] Wolters and Bickelhaupt[50] have examined a representative selection of hydrogen-, fluorine- and iodine-bonded model complexes, DX⋯A$^-$, using ZORA-BP86/TZ2P (where D = F and I; X = H, F and I; and A$^-$ = F$^-$, Cl$^-$, Br$^-$ and I$^-$) and found the interaction energy to be between −80.6 and −14.5 kcal mol$^{-1}$ (for IH⋯F$^-$ and IF⋯Cl$^-$, respectively). Intermolecular interactions of similar strength have been reported for halogen bonded systems, such as −N–X$^+$⋯$^-$O–N$^+$, wherein an oxygen atom served as an unusual halogen bond acceptor.[49] The data in Table 1 show a wide range of strengths of the ion-molecule interactions, which can be classified as ultra-strong in some cases (29-31), very strong (8-9,15-17), strong (1-4, 6-7,18-21,23, 26-28, 32), moderate (11, 13, 24), or weak (5, 14, 22, 25, 33-34), based on the degree of interaction

between the anion and the molecule, and the recommendation provided in an earlier study.[48] There is no obvious relationship notable between the $\Delta E$(BSSE) and the intermolecular distances within the entire series of complexes examined.

Our results indicate that the O atoms in molecules such as $X_2O$ (X = F, Cl, Br) and NCFO have tremendous potential to form unusually strongly bound complexes, driven by chalcogen bonding. This is especially when the electrophile on the O atom in molecules is exposed to anions.

**3.4 Second-order donor-acceptor charge transfer delocalization from NBO analyses**

Consistent with the work of others,[11] the results of the second-order perturbative estimates of donor-acceptor (bond-antibond) interactions obtained from an NBO analysis suggests that several complexes of Fig. 1 are the result of significant charge transfer (CT) delocalization between the σ* anti-bonding orbital of the chalcogen bond donor fragment and the lone-pair bonding orbitals associated with the bases A⁻. For example, the complexes $F_2O_2$···A⁻ (A = F, Cl, Br) are predominately due to the result of the $n$(A) → σ*(O–O) CT delocalization, with $E^{(2)}$ of 112.9, 159.9 and 198.44 kcal mol⁻¹ for $F_2O_2$···F⁻, $F_2O_2$···Cl⁻, and $F_2O_2$···Br⁻, respectively, where $n$ represents the lone-pair bonding orbital(s) of the anionic species. Similarly, the complexes $F_2O$···F⁻, $F_2O$···Cl⁻, and $F_2O$···Br⁻ are the results of the $n$(A) → σ*(O–F) CT, with $E^{(2)}$ of 184.4, 141.9 and 127.9 kcal mol⁻¹, respectively.

The complex $Cl_2O$···F⁻ is due to the combined effect of CT delocalizations such as $n(2)$(F) → σ*(O–Cl1)/σ*(O–Cl2) ($E^{(2)}$ = 8.3/1.0 kcal mol⁻¹), $n(3)$(F) → σ*(O–Cl1)/ σ*(O–Cl2) ($E^{(2)}$ = 0.3/1.8 kcal mol⁻¹) and $n(4)$(F) → σ*(O–Cl1)/ σ*(O–Cl2) ($E^{(2)}$ = 139.2/5.5 kcal mol⁻¹). For the complex $Cl_2O$···Cl⁻, the $E^{(2)}$ for the CT interactions, *viz.* $n(1)$(Cl) → σ*(O–Cl1)/ σ*(O–Cl2) and $n(3)$(Cl) → σ*(O–Cl1)/σ*(O–Cl2) are 2.3/2.3 and 2.3/2.3 kcal mol⁻¹, respectively. The corresponding CT delocalizations responsible for $Cl_2O$···Br⁻ are $n(1)$(Br) → σ*(O–Cl1)/σ*(O–Cl2) and $n(3)$(Br) → σ*(O–Cl1)/σ*(O–Cl2), with $E^{(2)}$ of 2.2 and 2.1 kcal mol⁻¹, respectively. These results indicate that several lone-pair bonding orbitals on the halide ion X$^-$ facilitate CT interactions with the anti-bonding σ* orbitals of the C–X bonds, providing evidence of the formation of chalcogen bonding in these complexes. The manyfold interaction topologies between the donor and acceptor orbitals are apparently due to the non-linear nature of the F⋯O, Cl⋯O and Br⋯O interactions in these complexes, respectively (∠F⋯O–Cl = 149.9° in Cl$_2$O⋯F$^-$, ∠Cl⋯O–Cl = 120.0° in Cl$_2$O⋯Cl$^-$, ∠Br⋯O–Cl = 118.5° in Cl$_2$O⋯Br$^-$).

The angle of interaction in the complexes Br$_2$O⋯F$^-$, Br$_2$O⋯Cl$^-$, and Br$_2$O⋯Br$^-$ is also significantly non-linear. For instance, ∠F⋯O–Br, ∠Cl⋯O–Br and ∠Br⋯O–Br are 124.0, 121.2 and 120.0° in the respective complexes, resulting in pseudo-windmill type geometries between the O and X atoms. There is, therefore, an expectation of orbital interaction in these complexes that could be associated with the two σ* anti-bonding orbitals of the two Br–O bonds in the OBr$_2$ moiety and the lone-pair bonding orbitals of the halide anions. Indeed, the second order analysis suggests such a possibility; the Br$_2$O⋯F$^-$ complex is the result of (*n(4)*(F) → σ*(O–Br1)) and (*n(4)*(F) → σ*(O–Br2)) delocalizations, each with an $E^{(2)}$ of 61.2 kcal mol$^{-1}$. The strongest orbital interaction occurs between *n(4)* on F and each of two σ*-orbitals of the O–Br bond; $E^{(2)}$ for analogous CT interactions involving the lone-pair bonding orbitals 2 and 3 on F and σ*(O–Br) were 5.4 and 2.0 kcal mol$^{-1}$, respectively. Similarly, for the complex Br$_2$O⋯Cl$^-$, the CT interactions were (*n(1)*(Cl) → σ*(O–Br1)/σ*(O–Br2) and (*n(3)*(Cl) → σ*(O–Br1)/σ*(O–Br2), each with $E^{(2)}$ of 2.4/1.9 kcal mol$^{-1}$. For the complex Br$_2$O⋯Br$^-$ the corresponding CT delocalizations were (*n(1)*(Br) → σ*(O–Br1)/σ*(O–Br2) and (*n(3)*(Br) → σ*(O–Br1)/σ*(O–Br2), each with an $E^{(2)}$ of 2.4/1.8 kcal mol$^{-1}$.

Along similar lines, the complexes of OX$_2$ (X = Cl, Br) and the O end of OCN$^-$ (22, 33) are described by *n*(O) → σ*(O–X) CT delocalizations. The $E^{(2)}$ for these are 1.76 and 2.36 kcal mol$^{-1}$, respectively. The trend in $E^{(2)}$ is consistent with the trend in Δ$E$(BSSE) of these complexes. The O···Br secondary interaction in Br$_2$O···OCN$^-$ is characterized by an $E^{(2)}$ of 0.38 kcal mol$^{-1}$.

The complex, F$_2$O$_2$···NC$^-$, is a result of the $\pi$(C≡N) → $\sigma$*(O–O) and $n$(N) → $\sigma$*(O–O) CT delocalizations, with $E^{(2)}$ of 0.63 and 1.20 kcal mol$^{-1}$, respectively. Although the former provides evidence of the involvement of a $\pi$···$\sigma$ interaction, the latter supports the formation of lone-pair—$\sigma$* type chalcogen bonding. In addition, and because the O atom in O$_2$F$_2$ is not entirely positive, the lone-pair orbital of O facilitates back-bonding CT interactions with the $\sigma$* orbital of the C≡N fragment (viz. $n$(O) → $\sigma$*(C≡N)), with $E^{(2)}$ of 1.42 kcal mol$^{-1}$. Similarly, the complex (CN)$_2$O···O(NO$_2$)$^-$ features CT delocalizations of $n$(O) (NO$_3$$^-$) → $\sigma$*(O–C) ((CN)$_2$O) ($E^{(2)}$ = 0.9 kcal mol$^{-1}$) and $n$(O) (NO$_3$$^-$) → $\pi$*(C≡N) ($E^{(2)}$ = 6.3 kcal mol$^{-1}$), showing that in addition to the formation of the lone-pair—$\sigma$* chalcogen bond, the $\sigma$—$\pi$* interaction plays a crucial role in determining the equilibrium geometry of the complex. Most of the complexes discussed above do involve several other weak CT interactions; detailed discussion of these is beyond the scope of this study.

QTAIM's bond path topologies discussed already above revealed that one of the two O···X interactions in the complexes of OX$_2$ with Br$_3$$^-$ (14, 25 and 34 of Fig. 1) is marginally stronger than the other. The CT delocalizations corresponding to these two interactions in F$_2$O···Br$_3$$^-$ are described by $n$(terminal Br) → $\sigma$*(O–F2) and $n$(middle Br) → $\sigma$*(O–F3), , with $E^{(2)}$ of 1.06 and 0.86 kcal mol$^{-1}$, respectively. The analogous delocalizations in Cl$_2$O···Br$_3$$^-$ were $n$(terminal Br) → $\sigma$*(O–Cl2) and $n$(middle Br) → $\sigma$*(O–Cl3), with $E^{(2)}$ of 0.32 and 1.12 kcal mol$^{-1}$, respectively. Similarly, these are $n$(terminal Br) → $\sigma$*(O–Br2) and $n$(middle Br) → $\sigma$*(O–Br3) for Br$_2$O···Br$_3$$^-$, with $E^{(2)}$ of 0.73 and 1.58 kcal mol$^{-1}$, respectively.

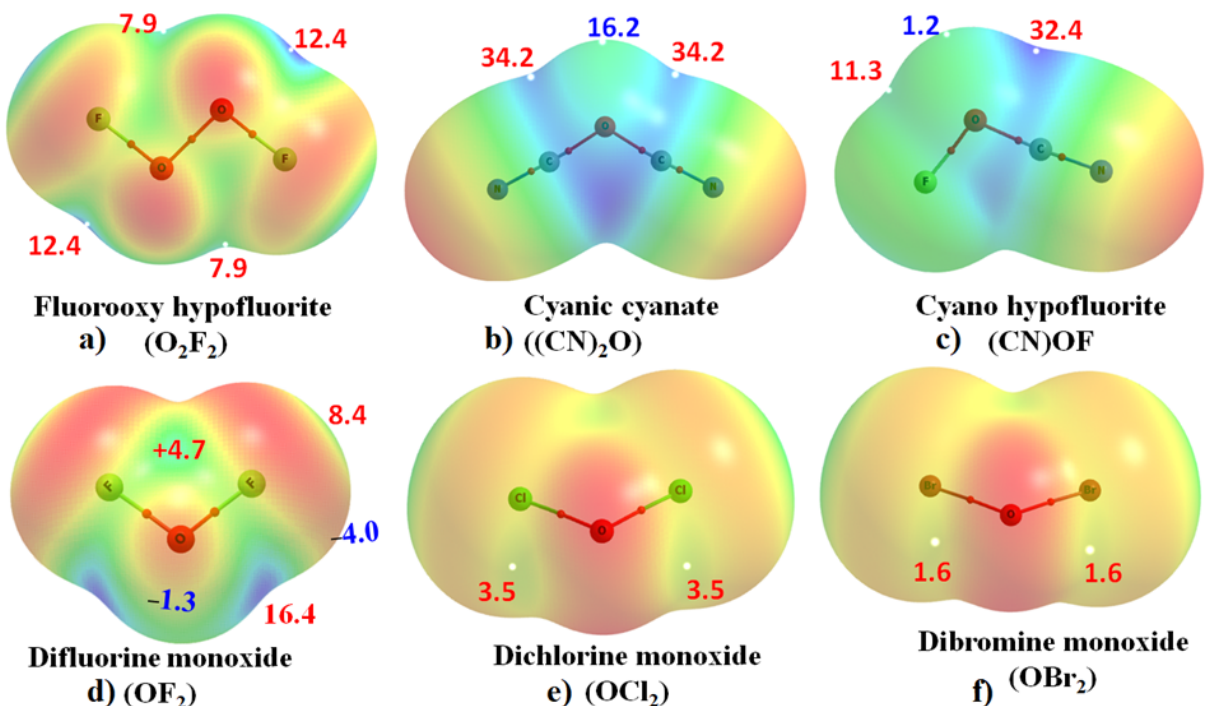

**Fig. 2**: 0.001 au isodensity envelope mapped molecular electrostatic surface potential for the O-containing monomers, obtained using MP2/aug-cc-pVTZ. Selected local maxima and minima of potential ($V_{S,max}$ and $V_{S,min}$, respectively) are shown in red and blue, respectively. QTAIM molecular graphs are shown for each monomer

### 3.5 Analysis of the MESP model results

The results of the MESP model shown in Fig. 2. The electrophilic character and the extent of electrophilicity of the O atom in the monomers can be rationalized by the positive sign and magnitude of $V_{s,max}$ on O, respectively. For example, the $V_{s,max}$ values were calculated to be 34.2, 32.4, 16.4, 3.5, 1.6 and 12.4 kcal mol$^{-1}$ on the C–O, F–O, F–O, Cl–O, Br–O and O–O bond extensions in (CN)$_2$O, NCFO, F$_2$O, Cl$_2$O, Br$_2$O, and F$_2$O$_2$, respectively (Fig. 2). These are all positive, and hence represent the O-centered σ-holes on respective monomers. Evidently, the intermolecular association between the interacting monomers shown in Fig. 1 can be rationalized by Coulomb's law. For instance, the positive site on the O atom along the X–O (X = O, F, Cl, Br, CN) bond extensions in F$_2$O$_2$, X$_2$O (X = F, Cl, Br), NCFO, and (CN)$_2$O attracts the electron rich anion species X$^-$ (X = F, Cl, Br), SCN$^-$, OCN$^-$, NO$_3^-$ and CN$^-$ by Coulombic interaction, leading to the formation of the ion-molecule complexes of Fig. 1. Similarly, the O atom along the F–O extension in NCFO has a positive $V_{S,max}$ of 32.4 kcal mol$^{-1}$ and is responsible for the formation of very strong complexes with Cl$^-$ and Br$^-$ (($\Delta E$(BSSE) for NCFO⋯Cl$^-$ (16) and NCFO⋯Br$^-$ (17) were found to be –48.74 and –51.46 kcal mol$^{-1}$, respectively).

On the other hand, the O atom on the F–O extension in F$_2$O has a relatively small positive $V_{S,max}$ of 16.4 kcal mol$^{-1}$, yet it is responsible for the formation of yet more stable complexes with the same anions (($\Delta E$(BSSE) for F$_2$O⋯Cl$^-$ (8) and F$_2$O⋯Br$^-$ (9) were –54.10 and –54.93 kcal mol$^{-1}$, respectively). The complexes of F$_2$O$_2$ with X$^-$ are comparatively weaker than those of Cl$_2$O⋯X$^-$ and Br$_2$O⋯X$^-$ (X = F, Cl, Br) (see Table 1 for $\Delta E$(BSSE) values). The above-mentioned preference in the stabilization energy is also unusual since $V_{S,max}$ on the O atom in F$_2$O$_2$ ($V_{S,max}$ = 12.4 kcal mol$^{-1}$) is larger than that in Cl$_2$O ($V_{S,max}$ = 3.5 kcal mol$^{-1}$) and in Br$_2$O ($V_{S,max}$ = 1.6 kcal mol$^{-1}$).

The latter results unequivocally demonstrate that the local most $V_{S,max}$ values associated with O's σ-hole in isolated monomers cannot be used as a universal descriptor of intermolecular interactions, or a measure of complex stability. This is arguably because the dependence between $V_{S,max}$ and $\Delta E$ has previously been demonstrated to be linear in some systems,[22,84-86] but observed not to be so in the complexes reported in this study. This suggests that the Coulombic view of explaining the origin of intermolecular interactions in the complexes examined in this study is incomplete, and such a view is not applicable to all systems as is not universal.[30,31,87,88] The potential limitations of the MESP model has been discussed;[17] it fails to explain Type-I halogen bonding and Type-III halogen-centered noncovalent interactions in complex systems. These interactions are the consequence of attraction between bonded atoms that have similar electrostatic potential; the former unusually does have a bent geometry and the latter is linear or quasi-linear. Another limitation of the model is that it fails to provide any insight into the origins of the attraction persisting between the interacting monomers of complexes of Fig. 1, which can be revealed using NBO's second-order analysis (see above, 3.4). Clearly, the standalone use of the MESP model may lead to erroneous interpretations of the nature and origin of intermolecular interactions in systems where multi-fold interaction topologies of van der Waal type play a crucial role in determining the stability to the equilibrium geometries.[8,9,88]

### 3.5 DFT-SAPT energy decomposition analysis

To provide further insight into the energetic origin of the intermolecular interactions, we carried out DFT-SAPT based EDA for some selected complexes. The results are summarized in Table 2. As expected, there is clear a difference between the $\Delta E$(BSSE) and $E_{int}$(SAPT) values for each complex. This is obviously due to the difference in the level of electron-electron correlation effects accounted for at each level of theory (MP2/aug-cc-VTZ and SAPT/aug-cc-pVDZ). Despite this difference, the SAPT analysis suggests that the electrostatic and polarization components of energy ($E_{eles}$ and $E_{ind}$, respectively) are very large and negative, contributing attractively to the overall interacting energy. This could be taken to mean that Columbic interactions play a vital role in stabilizing the equilibrium geometries of the ion-molecule complexes.

**Table 2.** DFT-SAPT decomposed energy components (kcal mol$^{-1}$) of some selected O-bonded complexes in their respective ground electronic states, obtained using the MP2/aug-cc-pVTZ equilibrium geometries. The BSSE-corrected MP2 binding energies are included for comparison.[a]

| No. | Complex | $E_{eles}$ | $E_{rep}$ | $E_{ind}$ | $E_{disp}$ | $E_{int}$(SAPT) | $\Delta E$(BSSE)(MP2) |
|---|---|---|---|---|---|---|---|
| 1 | $F_2O_2\cdots F^-$ | -38.58 | 71.86 | -50.17 | -8.06 | -24.94 | -22.23 |
| 2 | $F_2O_2\cdots Cl^-$ | -36.27 | 73.85 | -48.62 | -10.81 | -21.86 | -22.78 |
| 3 | $F_2O_2\cdots Br^-$ | -37.19 | 76.26 | -51.69 | -11.45 | -24.06 | -26.14 |
| 4 | $F_2O_2\cdots NC^-$ | -5.10 | 7.39 | -4.19 | -2.78 | -4.68 | -3.82 |
| 5 | $(CN)_2O\cdots NO_3^-$ | -25.02 | 21.89 | -9.61 | -7.69 | -20.43 | -18.36 |
| 6 | $F_2O\cdots F^-$ | -60.67 | 100.45 | -71.49 | -9.87 | -41.58 | -37.74 |
| 7 | $F_2O\cdots Cl^-$ | -72.21 | 146.91 | -109.29 | -17.09 | -51.69 | -54.1 |
| 8 | $F_2O\cdots Br^-$ | -68.20 | 138.39 | -103.18 | -16.86 | -49.85 | -54.93 |
| 9 | $F_2O\cdots NCO^-$ | -51.21 | 95.91 | -62.44 | -13.72 | -31.46 | -31.84 |
| 10 | $F_2O\cdots NCO^-$ | -10.08 | 13.32 | -5.71 | -3.74 | -6.21 | -6.08 |
| 11 | $F_2O\cdots SCN^-$ | -53.73 | 104.24 | -58.67 | -15.21 | -53.73 | -40.90 |
| 12 | $F_2O\cdots NCS^-$ | -5.58 | 6.89 | -2.45 | -3.73 | -4.86 | -4.89 |
| 13 | $Cl_2O\cdots OCN^-$ | -1.34 | 7.56 | -4.11 | -3.78 | -1.68 | -2.30 |
| 14 | $NCFO\cdots F^-$ | -69.84 | 95.26 | -71.42 | -10.54 | -56.54 | -50.86 |
| 15 | $NCFO\cdots Cl^-$ | -61.05 | 97.84 | -73.93 | -13.81 | -50.95 | -48.74 |
| 16 | $NCFO\cdots Br^-$ | -59.64 | 98.03 | -76.27 | -14.26 | -52.14 | -51.46 |

[a] $E_{int}$(SAPT) = $E_{eles}$ + $E_{exch}$ + $E_{ind}$ + $E_{disp}$

The most dominant energy term competing with $E_{eles}$ and $E_{ind}$ is the exchange term, $E_{exch}$, which is as large as 146.91, 138.39 and 100.45 kcal mol$^{-1}$ for $F_2O\cdots Cl^-$, $F_2O\cdots Br^-$ and $F_2O\cdots F^-$, respectively. The sum of the terms $E_{eles}$, $E_{ind}$ and $E_{exch}$ results in negative values for most of the complexes, showing that the stability of these complexes may be explained at the Hartree-Fock level. The inclusion of the dispersion part of the interaction energy, $E_{disp}$, to the sum ($E_{eles}$ + $E_{ind}$ + $E_{exch}$) leads $E_{int}$(SAPT) to be somewhat comparable to the $\Delta E$(BSSE) for each complex. This shows that the geometric stability of each ion-molecule complex system is the result of a delicate balance between all the four energy components. Clearly, one should not disregard any particular interaction energy term and focus on another in order to advance the argument that the stability of the interacting members in a complex is purely Columbic in orgin.[1,89] We have previously shown that such an argument can be misleading since it alone cannot explain the stability of the F⋯F interactions found in the complexes of $(C_6H_nF_m)_2$ (n, m = 0–6),[30] as well as of the X···X and X···O weakly bound interactions in many other complex systems.[8,9,29,31] A notable example in this study is the complex $Cl_2O···OCN^-$ in which the dispersion term dominates and the sum ($E_{eles}$ + $E_{ind}$ + $E_{exch}$) is positive.

## 4. Conclusion

This study examined a series of 34 molecule-anion complexes and has shown that the O-atom in molecules such as $OX_2$ (X = F, Cl, Br), $O_2F_2$, OFCN, and $(CN)_2O$ does indeed conceive a positive σ-hole, described by the presence of a positive electrostatic potential along the R–O bond extension. These σ-holes have a significant ability to promote the formation of O-centered chalcogen bonds with various anions, which interactions may or may not be strictly directional. The interaction energies of the resulting ion-molecule complexes could be classified as weak, moderate, strong, very strong, or ultra-strong. The nature of stability is clearly controlled by the nucleophilicity of the anionic bases, and the extent of their attractive engagements with the electrophilic site the O atom in the molecules studied. The results of the MESP model explained the intermolecular interactions in the complexes as of coulombic origin (*viz.* a negative site attracts a positive one). However, the magnitude of $V_{S,max}$ on the O atom of the monomers is shown to be inadequate to explain the trend in the $\Delta E$(BSSE) values found for this series of complexes. This result led to the conclusion that $V_{S,max}$ cannot be used as a universal descriptor of the energy of intermolecular interactions. The NBO's second-order analysis results showed that the origin of the chalcogen bonding is predominantly of $n \rightarrow \sigma^*$ type charge transfer delocalization, and the magnitude of $E^{(2)}$ associated these delocalizations were unusually large for some complexes. The results of QTAIM showed that the great majority of the interactions, of whatever flavor, in these molecule-anion complexes, are not only ionic but also contain appreciable covalent character; this is consistent with the large polarization and dispersion energies computed for these complexes.

**Author Contributions:** Conceptualization, problem design, computation, data analysis, chemical system drawing, interpretation, paper writing and editing: PRV and AV; Editing and Review: PRV, AV and HMM.

**Funding:** This research received no external funding. Accordingly, the funders had no role in the design, conceptualization and investigation of the study; in the handling, analyses, or interpretation of data; in the writing of the manuscript; or in our decision to publish the results.

**Acknowledgments:** This work was entirely conducted using the various facilities provided by the University of Tokyo and the computing center of Institute of Molecular Sciences (Okazaki). P.R.V. and A.V. thank Prof. K. Yamashita for support. A.V. is currently affiliated with AIST. H.M.M. acknowledges the financial support of the University of the Witwatersrand.

**Conflicts of Interest:** The authors declare no conflicts of interest.

**Reference**
1. Politzer, P.; Murray, J. S. Crystals 2017, 7, 212, doi:210.3390/cryst7070212.
2. Politzer, P.; Murray, J. S.; Clark, T.; Resnati, G. Phys Chem Chem Phys 2017, 19, 32166-32178.
3. Beno, B. R.; Yeung, K.-S.; Bartberger, M. D.; Pennington, L. D.; Meanwell, N. A. J Med Chem 2015, 58, 4383-4438.
4. Bauzá, A.; Mooibroek, T. J.; Frontera, A. ChemPhysChem 2015, 16, 2496-2517.
5. Mukherjee, A. J.; Zade, S. S.; Singh, H. B.; Sunoj, R. B. Chem Rev 2010, 110, 4357-4416.
6. Huang, H.; Yang, L.; Facchetti, A.; Marks, T. J. Chem Rev 2017, 117, 10291-10318.
7. Benz, S.; Mareda, J.; Besnard, C.; Sakai, N.; Matile, S. Chem Sci 2017, 8, 8164-8169.
8. Varadwaj, P. R. Molecules 2019, 24, 3166.
9. Varadwaj, P. R.; Varadwaj, A.; Marques, H. M.; MacDougall, P. J. Phys Chem Chem Phys 2019, 21, 19969-19986.
10. Vogel, L.; Wonner, P.; Huber, S. M. Angew Chem Int Edn 2019, 58, 1880-1891.
11. Pascoe, D. J.; Ling, K. B.; Cockroft, S. L. J Am Chem Soc 2017, 139, 15160-15167.
12. Scilabra, P.; Terraneo, G.; Resnati, G. Acc Chem Res 2019, 52, 1313-1324.
13. Dong, W.; Li, Q.; Scheiner, S. Molecules 2018, 23, 1681.
14. Guo, X.; An, X. L.; Li, Q. Z. J Phys Chem A 2015, 119, 3518-3527.
15. Adhikari, U.; Scheiner, S. J Phys Chem A 2014, 118, 3183-3192.
16. Clark, T.; Hennemann, M.; Murray, J. S.; Politzer, P. J Mol Model 2007, 13, 291-296.
17. Varadwaj, P. R.; Varadwaj, A.; Marques, H. M. Inorganics 2019, 7, 40.
18. Navarro-García, E.; Galmés, B.; Velasco, M. D.; Frontera, A.; Caballero, A. Chem Eur J 2020, 26, 4706-4713.

19. Fellowes, T.; Harris, B. L.; White, J. M. Chem Commun 2020, 56, 3313-3316.
20. Silvi, B.; Alikhani, E.; Ratajczak, H. J Mol Model 2020, 26, 62.
21. Stone, A. J. J Am Chem Soc 2013, 135, 7005-7009.
22. Politzer, P.; Murray, J. S.; Clark, T. Phys Chem Chem Phys 2010, 12, 7748-7757.
23. Mahmudov, K. T.; Kopylovich, M. N.; Guedes da Silva, M. F. C.; Pombeiro, A. J. L. Dalton Trans 2017, 46, 10121-10138.
24. Sarkhel, S.; Desiraju, G. R. Proteins: Structure, Function, and Bioinformatics 2004, 54, 247-259.
25. Schneider, H.-J. Acta Cryst 2018, B74, 322-324.
26. Steiner, T.; R. Desiraju, G. Chem Commun 1998, 891-892.
27. Dimitrijević, E.; Kvak, O.; Taylor, M. S. Chem Commun 2010, 46, 9025-9027.
28. Varadwaj, A.; Varadwaj, P. R. Chem Eur J 2012, 18, 15345-15360.
29. Varadwaj, A.; Marques, H. M.; Varadwaj, P. R. J Comp Chem 2019, 40, 1836-1860.
30. Varadwaj, P. R.; Varadwaj, A.; Marques, H. M.; Yamashita, K. Computation 2018, 6, 51.
31. Varadwaj, A.; Marques, H. M.; Varadwaj, P. R. Molecules 2019, 24, 379.
32. Varadwaj, P. R.; Varadwaj, A.; Marques, H. M. Sci Rep 2019, 9, 10650.
33. Stephen, G. D.; Haoran, S. Curr Top Med Chem 2006, 6, 1473-1482.
34. Wang, J.; Guo, Z.; Fu, Y.; Wu, Z.; Huang, C.; Zheng, C.; Shar, P. A.; Wang, Z.; Xiao, W.; Wang, Y. Briefings Bioinform 2016, 18, 321-332.
35. Evans, E.; Ritchie, K. Biophys J 1999, 76, 2439-2447.
36. Hutchins, K. M. Roy Soc Open Sci 2018, 5, 180564.
37. Varadwaj, P. R. Polymers 2020, 12, 1054.
38. Turner, D., Ed. Crystals. Special Issue: Crystal Engineering Involving Weak Bonds; MDPI Basel, Switzerland, 2014.
39. Jungbauer, S. H.; Bulfield, D.; Kniep, F.; Lehmann, C. W.; Herdtweck, E.; Huber, S. M. J Am Chem Soc 2014, 136, 16740-16743.
40. Głowacki, E. D.; Irimia-Vladu, M.; Bauer, S.; Sariciftci, N. S. J Mat Chem B 2013, 1, 3742-3753.
41. Pollard, T. D.; Earnshaw, W. C.; Lippincott-Schwartz, J.; Johnson, G. T. Cell Biology; Elsevier, Inc.: Philadelphia, PA, 2017.
42. Ege, S. Organic Chemistry: Structure and Reactivity; Houghton Mifflin College: Boston, MA, 2003.
43. Larson, J. W.; McMahon, T. B. Inorg Chem 1984, 23, 2029-2033.
44. Emsley, J. Chem Soc Rev 1980, 9, 91-124.
45. Jeffrey, G. A. An Introduction to Hydrogen Bonding; Oxford University Press: Oxford, UK, 1997.
46. Desiraju, G. R.; Steiner, T. The weak hydrogen bond in structural chemistry and biology (International Union of Crystallography Monographs on Crystallography, 9); Oxford University Press: Oxford and New York, 1999.

47. D'Oria, E.; Novoa, J. J. J Phys Chem A 2011, 115, 13114-13123.
48. Varadwaj, A.; Varadwaj, P. R.; Yamashita, K. J Comp Chem 2017, 38, 2802–2818
49. Puttreddy, R.; Jurcek, O.; Bhowmik, S.; Makela, T.; Rissanen, K. Chem Commun 2016, 52, 2338-2341.
50. Wolters, L. P.; Bickelhaupt, F. M. ChemistryOpen 2012, 1, 96-105.
51. Varadwaj, A.; Varadwaj, P. R.; Yamashita, K. ChemSusChem 2018, 11, 449-463.
52. Alkorta, I.; Elguero, J. New J Chem 2018, 42, 13889-13898.
53. Varadwaj, P. R.; Varadwaj, A.; Marques, H. M.; Yamashita, K. Scientific Reports 2019, 9, 50.
54. Groom, C. R.; Bruno, I. J.; Lightfoot, M. P.; Ward, S. C. Acta Cryst 2016, B72, 171-179.
55. Kumar, V.; Leroy, C.; Bryce, D. L. CrystEngComm 2018, 20, 6406-6411.
56. Riel, A. M. S.; Huynh, H.-T.; Jeannin, O.; Berryman, O.; Fourmigué, M. Cryst Growth & Design 2019, 19, 1418-1425.
57. Lee, L. M.; Corless, V.; Luu, H.; He, A.; Jenkins, H.; Britten, J. F.; Adam Pani, F.; Vargas-Baca, I. Dalton Trans 2019, 48, 12541-12548.
58. Galmés, B.; Juan-Bals, A.; Frontera, A.; Resnati, G. Chem Eur J 2020, 26, 4599-4606.
59. Alkorta, I.; Elguero, J.; Frontera, A. Crystals 2020, 10, 180.
60. Frisch, M. J.; Head-Gordon, M.; Pople, J. A. Chem Phys Lett 1990, 166, 275-280.
61. SAPT: Symmetry Adapted Perturbation Theory, http://www.psicode.org/psi4manual/master/sapt.html., accessed September 19, 2019.
62. Turney, J. M.; Simmonett, A. C.; Parrish, R. M.; Hohenstein, E. G.; Evangelista, F.; Fermann, J. T.; Mintz, B. J.; Burns, L. A.; Wilke, J. J.; Abrams, M. L.; Russ, N. J.; Leininger, M. L.; Janssen, C. L.; Seidl, E. T.; Allen, W. D.; Schaefer, H. F.; King, R. A.; Valeev, E. F.; Sherrill, C. D.; Crawford, T. D. WIREs Comput Mol Sci 2012, 2, 556–565.
63. Weinhold, F.; Landis, C. R. Discovering Chemistry with Natural Bond Orbitals; John Wiley & Sons, Inc.: Hoboken, NJ, USA, 2012.
64. Bader, R. F. W.; Nguyen-Dang, T. T. In Advances in Quantum Chemistry; Löwdin, P.-O., Ed.; Academic Press: New York, NY, USA, 1981, p 63–124.
65. Bhattarai, S.; Sutradhar, D.; Chandra, A. K.; Zeegers-Huyskens, T. Molecules 2020, 25, 416.
66. Varadwaj, P. R.; Varadwaj, A.; Marques, H. M. Crystals 2020, 10, 146.
67. Bauzá, A.; Alkorta, I.; Frontera, A.; Elguero, J. J Chem Theory Comput 2013, 9, 5201-5210.
68. M. J. Frisch; G. W. Trucks; H. B. Schlegel; G. E. Scuseria; M. A. Robb; J. R. Cheeseman; G. Scalmani; V. Barone; B. Mennucci; G. A. Petersson; H. Nakatsuji; M. Caricato; X. Li; H. P. Hratchian; A. F. Izmaylov; J. Bloino; G. Zheng; J. L. Sonnenberg; M. Hada; M. Ehara; K. Toyota; R. Fukuda; J. Hasegawa; M. Ishida; T. Nakajima; Y.

Honda; O. Kitao; H. Nakai; T. Vreven; J. A. Montgomery, J.; J. E. Peralta; F. Ogliaro; M. Bearpark; J. J. Heyd; E. Brothers; K. N. Kudin; V. N. Staroverov; R. Kobayashi; J. Normand; K. Raghavachari; A. Rendell; J. C. Burant; S. S. Iyengar; J. Tomasi; M. Cossi; N. Rega; J. M. Millam; M. Klene; J. E. Knox; J. B. Cross; V. Bakken; C. Adamo; J. Jaramillo; R. Gomperts; R. E. Stratmann; O. Yazyev; A. J. Austin; R. Cammi; C. Pomelli; J. W. Ochterski; R. L. Martin; K. Morokuma; V. G. Zakrzewski; G. A. Voth; P. Salvador; J. J. Dannenberg; S. Dapprich; A. D. Daniels; Ö. Farkas; J. B. Foresman; J. V. Ortiz; Cioslowski, J.; Fox, D. J.; Gaussian, Inc.: Wallinford, CT, 2009.

69. Murray, J. S.; Politzer, P. WIREs Comput Mol Sci 2011, 1, 153–163.
70. Lu, T.; Chen, F. J Comput Chem 2012, 33, 580-592.
71. Keith, T. A.; TK Gristmill Software, http://aim.tkgristmill.com: Overland Park, KS, 2016.
72. Boys, S. F.; Bernardi, F. Mol Phys 1970, 19, 553-566.
73. Bader, R. F. Atoms in Molecules: A Quantum Theory; Oxford University Press: Oxford, 1990.
74. Matta, C. F.; Boyd, R. J. T. The Quantum Theory of Atoms in Molecules; Wiley-VCH: Weinheim, 2007.
75. Poater, J.; Duran, M.; Solà, M.; Silvi, B. Chem Rev 2005, 105, 3911-3947.
76. Fradera, X.; Austen, M. A.; Bader, R. F. W. J Phys Chem A 1999, 103, 304-314.
77. Aakeroy, C. B.; Bryce, D. L.; Desiraju, R. G.; Frontera, A.; Legon, A. C.; Nicotra, F.; Rissanen, K.; Scheiner, S.; Terraneo, G.; Metrangolo, P.; Resnati, G. Pure and Applied Chem 2019, 91, 1889–1892.
78. Cremer, D.; Kraka, E. Croat Chem Acta 1984, 57, 1259-1281.
79. Berryman, V. E. J.; Whalley, Z. J.; Shephard, J. J.; Ochiai, T.; Price, A. N.; Arnold, P. L.; Parsons, S.; Kaltsoyannis, N. Dalton Trans 2019, 48, 2939-2947.
80. Cabeza, J. A.; Van der Maelen, J. F.; García-Granda, S. Organometallics 2009, 28, 3666-3672.
81. Macchi, P.; Garlaschelli, L.; Sironi, A. J Am Chem Soc 2002, 124, 14173-14184.
82. Tiana, D.; Francisco, E.; Macchi, P.; Sironi, A.; Martín Pendás, A. J Phys Chem A 2015, 119, 2153-2160.
83. Pople, J. A. Faraday Discuss 1982, 73, 7-17.
84. Mohan, N.; Suresh, C. H. J Phys Chem A 2014, 118, 1697-1705.
85. Esrafili, M. D.; Ghanbari, M.; Mohammadian-Sabet, F. J Mol Model 2014, 20, 2436.
86. Politzer, P.; Murray, J. S. ChemPhysChem 2013, 14, 278-294.
87. Varadwaj, A.; Varadwaj, P. R.; Yamashita, K. J Comput Chem 2018, 39, 343-350.
88. Varadwaj, A.; Varadwaj, P. R.; Marques, H. M.; Yamashita, K. ChemPhysChem 2018, 19, 1486–1499.
89. Politzer, P.; Murray, J. S. Crystals 2019, 9, 165.